# Terabit-class coherent communications enabled by an integrated photonics erbium doped amplifier

Di Che,[1,†,*] Stefano Grillanda,[1,†] Yang Liu,[2] Zheru Qiu,[2] Xinru Ji,[2] Gregory Raybon,[1] Xi Chen,[1] Kwangwoong Kim,[1] Tobias J. Kippenberg,[2] Andrea Blanco-Redondo,[1,3]

[1]Nokia Bell Labs, Murray Hill, NJ 07974, USA
[2]Institute of Physics, Swiss Federal Institute of Technology Lausanne (EPFL), CH-1015 Lausanne, Switzerland
[3]CREOL, The College of Optics and Photonics, University of Central Florida, Orlando, FL 32816, USA
[†]Equal contributors
[*]Corresponding author. Email: di.che@nokia-bell-labs.com

**Coherent technologies have revolutionized optical communications, driving the capacity per fiber to multi-terabit per second (Tb/s) in combination with wavelength division multiplexing (WDM). With an ever-increasing deployment density of coherent systems, the demand for highly integrated WDM coherent transceivers has been rising. While tremendous progress has been made on silicon photonics compatible high-speed modulation and photodetection on chip, a solution for monolithically integrable amplifier with high gain and output power remains a challenge. Recently, an erbium doped waveguide amplifier based on ultra-low loss silicon nitride waveguides has demonstrated gain and output power levels potentially suitable for Terabit class coherent communications. Here, we demonstrate a WDM coherent system enabled by this integrated photonic amplification solution. The system uses the waveguide amplifier as a booster amplifier of 16 WDM signals each carrying a net data rate of 1.6 Tb/s, achieving 25.6-Tb/s net capacity over 81-km fiber transmission. Our results highlight a fully integrated solution for highly parallel coherent transceivers including amplification, that has the potential to transform future optical communications.**

# Main

The revival of coherent technologies in the 2000s has driven optical communications to a multi-terabit per second (Tb/s) era over the past two decades [1]. Together with the early success in long-haul transmissions, coherent communications have gradually extended the application to a wide range of short-reach interconnects [2] with ever-increasing deployment volume and density. This adds stringent requirements on both the size and power consumption for coherent optics, making photonic integrated circuits (PICs) an indispensable solution [3]. However, while PICs have permeated many of the optical transceiver functionalities – modulators, detectors, passives – high power integrated photonic amplification has remained a missing piece.

Since its birth in the 1980s, Erbium doped fiber amplifiers (EDFAs) [4,5] have been the main force of fiber optic transmissions, that provides broadband gain over tens of nanometers to enable wavelength-division multiplexing (WDM), with low-noise figures approaching the quantum mechanical limit of 3 decibels (dB) and negligible nonlinear distortions (e.g. inter-channel crosstalk). Rare-earth ion doping of semiconductor waveguides can provide a similarly attractive avenue for amplification in photonic integrated platforms [6]. Pioneering efforts were made in the 1990s to implement erbium doped waveguide amplifiers (EDWAs) based on silica waveguides [7,8] and, more recently, in $Al_2O_3$ [9,10] and $TeO_2$ [11] semiconductor waveguides. However, the relatively high loss and the limited propagation lengths available in these platforms limited their output power to <1 mW, significantly below the power levels required in modern WDM transmission links. A recent breakthrough based on high-concentration ion-implantation of Erbium ions on ultra-low loss silicon nitride waveguides has shown to overcome these limitations [12]. Specifically, the work has shown both high-gain (>30 dB) and high output power (145 mW), on par with commercial EDFAs and surpassing state-of-the-art III-V heterogeneously integrated semiconductor optical amplifiers (SOAs) [13], which indicates the potential of this novel EDWA technology as a photonic-integrated amplification solution for coherent communications. However, the question remains whether the noise figure demonstrated on these EDWAs so far (7.1 dB) and the nonlinearities (e.g. Kerr nonlinearity) existing in silicon nitride waveguides allow their use in WDM coherent transmissions with broadband signaling at Terabit class.

Here, we successfully demonstrate Terabit-class coherent transmissions enabled by an EDWA based on silicon nitride photonic integrated circuits [14]. Specifically, we use the EDWA as a booster amplifier to support 16 WDM channels transmitting over 81-km single-mode fiber. Each channel carries a 170-GBaud signal on a WDM grid of 200 GHz, rendering a 25.6-Tb/s net capacity across a frequency span of 3.2 THz in C-band (i.e., the conventional band). The data rate goes far beyond previous EDWA-supported transmission experiments [15] and is on par with state-of-the-art lab demonstrations of C-band WDM transmission based on EDFAs [16]. These results highlight a solution for high power amplification in a small form factor at $mm^2$ scale, that could facilitate highly integrated coherent optics with optical amplifiers and multiple parallel transceivers on the same chip.

## Results

The EDWA is made by an Erbium (Er) ion doped silicon nitride ($Si_3N_4$) PIC with an Archimedean spiral layout that allows to pack waveguides up to meter-scale length on a small footprint of only a few $mm^2$. Figure 1a shows a top-view photograph of the EDWA chip used in our experiment, which hosts 15 Er-doped waveguides ranging from 5 to 50 cm. Figure 1b show a scanning electron microscope photograph of the EDWA cross section: it is formed by an air-cladded, Er-doped $Si_3N_4$ waveguide core with thickness of 0.7 μm and width of 2.1 μm, and has a silicon dioxide ($SiO_2$) under-cladding and a lateral cladding that minimizes the deformation under ion implantation. Note that no additional top cladding is added to the waveguide in order to allow direct implantation of Er-ions. The passive, low-loss $Si_3N_4$ waveguide was fabricated using the photonic damascene process [17] that can attain losses below 3dB/m. In order to maximize the overlap factor between the transverse electric (TE) optical mode and the Er-ion concentration, Er-ion implantation of the passive $Si_3N_4$ waveguides was performed in successive steps of different energy (~MeV-scale) and different ion fluences (~$10^{15}$ $cm^{-2}$ -scale). The waveguides after Er-ion implantation exhibit a peak Er-ion concentration of ~$3.25\times10^{20}$ $cm^{-3}$ (corresponding to 0.35% atomic concentration) with an overlap between optical mode intensity and Er-ions of about 50% (Fig. 1c). More details on the implantation process can be found in [12].

In the optical transmission experiments reported in this work, we used two different EDWAs as indicated in Fig. 1a, one (EDWA #1) with total length of 22.2 cm and footprint of ~1.8 mm$^2$, and the other (EDWA #2) with total length of 12.5 cm and footprint of ~1.0 mm$^2$. We measured the optical output power and gain provided by the EDWAs with an experimental setup shown in Fig. 2a, using a continuous-wave (CW) light at 1560 nm as an optical input. To maximize the amplification gain, we use both forward- and backward-propagating pumps at 1480 nm (additional details on the setup can be found in Methods). Figures 2(b-c) show, respectively, the output power and the gain of the two EDWAs measured with various input power. Note that the values of input/output power and gain were measured "off-chip", which include the fiber-to-chip coupling loss. For completeness, the secondary vertical axis of Figs. 2(b-c) also reports the "on-chip" output power and gain, where the coupling loss to the chip has been de-embedded. Each of the two 1480-nm pumps emitted about 24-dBm power in this measurement, which corresponded to 21.1-dB actual pump power to the chip considering 2.9 dB/facet of coupling loss. In these conditions, the two EDWAs have almost the same performance. In particular, the output power is about 12 dBm when the input power is -12 dBm and increases to more than 14.5 dBm when the input power is 2.8 dBm. Correspondingly, the gain varies from 12 dB up to 24 dB. For gain values higher than 20 dB (*i.e.*, on-chip net gain higher than about 26 dB), we have observed parasitic lasing due to the Fabry-Perot cavity created by the chip facets.

While EDWAs can be potentially implemented to long-haul fiber transmissions with multiple optical amplification stages, here we mainly focus on short-reach transmissions with a single span of fiber, which are in more urgent need of a small-form-factor optical amplifier to be co-integrated with transceivers. In a coherent communication system, the signal is modulated on both optical amplitude and phase for I/Q signaling (e.g., quadrature amplitude modulation, QAM), which requires an optical I/Q modulator to be biased at its null point. The optical power loss at the output of the modulator with respect to the CW input can be higher than 20 dB [2]. This makes the booster amplifier a critical component as well as the main limiting factor of the optical signal-to-noise ratio (OSNR) performance. We used the EDWA as a booster to amplify modulated signals. As a result, both the linear optical amplified spontaneous emission noise (ASE) and potential nonlinear noises generated during the propagation in the Er-doped waveguide would be overlapped in-band with the continuous-band signal and degrade the system performance. We considered two coherent application scenarios and designed two transmission experiments correspondingly. Experiment 1 was a single-channel transmission, for which the EDWA was supposed to be integrated inside a traditional single-wavelength coherent transmitter. Experiment 2 used one EDWA to amplify 16 dense WDM (DWDM) channels simultaneously, to highlight a promising highly parallel coherent solution that co-integrates a multi-wavelength light source like a frequency comb [18], wavelength (de-)multiplexers, parallel modulators, and EDWAs all together on the same chip.

**Single-channel transmission experiments.** The setup for the single-channel Experiment 1 (using EDWA #1 in Fig. 1a) is depicted in Fig. 3a. A single-polarization I/Q modulator was modulated by a probabilistically shaped (PS) [19] 64-QAM signal with an entropy of 5.7 bits/symbol and a symbol rate of 170 GBaud. The signal was pulse-shaped by a root-raised cosine (RRC) filter with a roll-off factor of 0.1, leading to approximately rectangular spectrum with 187-GHz bandwidth (Fig. 3c). The $Si_3N_4$ waveguide was optimized for transverse-electric (TE) polarization with a minimum background loss below 5 dB/m, and a slightly higher loss for transverse-magnetic (TM) polarization. Because of this, the EDWA was placed after the single-polarization I/Q modulator, followed by a polarization-division multiplexing (PDM) emulator. An integrated tunable laser assembly (ITLA) changed the transmitter channel frequency with a step size of 200 GHz aiming to test the EDWA performance at wavelengths across the entire C-band. Correspondingly, a second ITLA served as the local oscillator (LO) of a dual-polarization (DP) coherent receiver (additional details on the transmission setup can be found in Methods). For all wavelengths, the input power to the EDWA was -8 dBm, and the off-chip EDWA gain was kept to about 11 dB as shown in Fig. 3d by adjusting the pump power. We used offline digital signal processing (DSP) to process the coherently down-converted data. We chose normalized generalized mutual information (NGMI) as the system metric, and selected a soft-decision (SD) forward error correction (FEC) scheme with a code rate of 0.8402 and an NGMI threshold of 0.8798. The received signal can be decoded as error free if its pre-FEC NGMI is higher than the FEC threshold. The post-FEC net bit rate is 1.6 Tb/s for this 170-GBaud PDM signal. As shown in Fig. 3b, from 191.4 to 195.6 THz, all the wavelengths achieve the NGMI above the FEC threshold. To map our single-channel experiment to a real-world application, we also successfully transmit the signal over 10-km standard single-mode fiber (SSMF) using the 400ZR-defined frequency of 193.7 THz for grey coherent optics [2], with a received PS 64-QAM constellation shown in Fig. 3e.

**Multi-channel transmission experiments.** As the EDWA exhibited a lower noise figure at longer wavelengths [12], we chose 16 wavelengths starting from 191.4 to 194.4 THz with a grid of 200 GHz to demonstrate a DWDM Experiment 2, with the setup shown in Fig. 4a. The channel under test (CUT) used the same setup as the single-channel experiment shown in Fig. 3a, and the 15 side channels were generated by sending 15 wavelengths to a second I/Q modulator followed by a 1-km spool of standard-single-mode-fiber (SSMF) to decorrelate them. EDWA #2 in Fig. 1a served as a booster amplifier to amplify all the 16 channels. The on-chip forward pump power was set to 21.6 dBm and the backward was 21.9 dBm, which provided an on-chip gain of 16 dB and an off-chip gain of 10 dB. With -8 dBm signal power per wavelength (the same value as in the single-channel experiment), the total fiber input power to the EDWA was 4 dBm and the fiber-coupled output power was about 14 dBm. The ASE spectrum of the EDWA is shown in Fig. 4b when no input signal was injected, and the 16-channel WDM spectrum after the EDWA is shown in Fig. 4c. The DWDM signal was transmitted over 81-km Corning Vascade® EX2000 single-mode fiber and pre-amplified by a single-mode EDFA before coherent detection. This EDFA can be replaced by an EDWA that supports DP amplification in the future when it becomes available. At the receiver, the CUT was selected by a wavelength selective switch (WSS) with 200-GHz passband bandwidth for coherent detection. As shown in Fig. 4d, all the channels achieve the NGMI above the FEC threshold, leading to a 16-channel DWDM transmission with a net capacity of 25.6 Tb/s. Figure 4e shows example constellations of 170-GBaud PS 64-QAM signals at frequencies of 191.4, 192.4, 193.4 and 194.4 THz, respectively, after 81-km transmission.

## Discussion

We demonstrated the capability of the EDWA to support modern coherent communications at multi-Tb/s. Specifically, we achieved 25.6-Tb/s (16 × 1.6-Tb/s) net capacity coherent DWDM transmission over 81-km of fiber. A key enabler is the high output power of the EDWA that reaches an on-chip level of about 20dBm, on par with the commercially deployed EDFAs. As a reference, the coherent 400ZR implementation agreement [2] defines a 25.6-Tb/s fully loaded C-band application using 64 DWDM 400-Gb/s channels from 191.375 to 196.1 THz, which requires fiber launch power at a level of 20 dBm to support the ZR distance of 80 km. With an optimized chip-to-fiber coupling in the future, it would be possible to use a single EDWA at mm$^2$ scale to amplify a multi-Tb/s coherent DWDM signal across the C-band with >4-THz bandwidth.

With such high-power propagating through the Si$_3$N$_4$ waveguide, a big concern is whether the nonlinearities would become a major impairment of the transmission system. In earlier work [12], the EDWA was applied to the amplification of a broadband microcomb, for which nonlinearities like four-wave mixing may be hidden in the optical spectrum measurement because of the evenly spaced comb lines. In contrast, for a carrier-suppressed I/Q modulated signal with continuous spectrum, the nonlinear noise will be overlapped in-band with the signal and translate to system penalties. The demonstrated performance that supports both high symbol rate (170 GBaud) and high spectral efficiency per channel (net 9.48 bits/s/Hz) indicates the nonlinearities were not presented as a dominant limiting factor for this Terabit-class transmission.

The noise figure of the current EDWA setup is mainly limited by the fiber-to-chip coupling loss [12], which can be improved with optimized mode adapters placed at the chip facets. In DWDM Experiment 2, we chose 16 channels on the longer wavelength side of the spectrum with 3.2-THz span for slightly lower noise figure (about 7 dB on average [12]). While this does not cover the shorter wavelengths, the transmission experiment is expected to be straightforwardly extended to cover the full C-band. In fact, the uneven noise figure is also a common feature for commercial EDFAs. A gain flattening filter (GFF) is usually embedded inside an EDFA to equalize the gain across the amplifier's operating bandwidth. Moreover, for highly parallel point-to-point applications without wavelength routing, each wavelength channel can use rate-adaptation techniques (like entropy loading [20]) to match its transmission rate with the underlying OSNR. This will further extend the available bandwidth of an EDWA without the concern of the uneven noise figure or the additional insertion loss from the GFF.

Besides the booster amplifier after the signal modulation, the EDWA can be applied to a variety of positions in a transmission system, like a booster after the light source, a pre-amplifier ahead of the coherent receiver, or in-line amplifiers in multi-span fiber transmissions. Note that modern coherent transmissions commonly use PDM to double the spectral efficiency and enable the digital compensation of polarization mode dispersion. While the EDWA was demonstrated for single-polarization amplification in our experiments, it can be potentially developed for dual-polarization amplification by optimizing waveguide dimensions to minimize the polarization dependent loss.

Alternatively, polarization diversity may be realized by integrating a pair of EDWAs with polarization rotators and multiplexers on the same chip, a technique that has been demonstrated on the silicon nitride platform.

Besides the reported C-band WDM transmission, the co-doping of Er-ions with other rare-earth ions (such as ytterbium and thulium) can unlock the optical amplification in other wavelength regions in a mm$^2$-scale form factor, therefore supporting an ultra-wideband coherent transmission with hundreds of Terabit/s like what has been demonstrated with fiber amplifiers [21]. Moreover, such an active $Si_3N_4$ device is compatible with heterogeneous integration of other silicon-based components in a coherent transceiver like electro-optic modulators, photodiodes, polarization circuits, WDM components, etc. This enables a large-scale integration of an EDWA device with coherent transceivers on the same chip to offer an extremely high data-rate density in an order of Tb/s/mm$^2$. Consequently, EDWAs have great potential to replace the role of EDFAs and transform the deployment model of coherent transceivers in highly parallel coherent communications.

**ACKNOWLEDGMENTS** Y.L. was supported by the Swiss National Science Foundation under grant no. 221540 (BRIDGE PoC). This work was further supported by the EU Horizon Europe EIC transition programme under grant agreement No. 101113302 (MAGNIFY), which received funding by the Swiss State Secretariat for Education, Research and Innovation (SERI).

## Methods

**Evaluation setup of the EDWA chip.** The integrated EDWA was excited by pump lights centered around 1480 nm that were injected into the photonic chip in both forward and backward directions. The light was coupled to/from the chip using a pair of ultra-high numerical aperture fibers (UHNA 7) butt-coupled to the EDWA chip facets. The mode field diameter (MFD) of the UHNA fiber was about 3.2 $\mu$m and approached the MFD of the inverse tapers placed at the input and output of the EDWA chip. Index matching oil is used between the fibers and the EDWA chip. The fiber to chip coupling loss in the C-band was about 2.9 dB per chip facet with this coupling scheme. We chose 1480 nm pumping due to the lower fiber-to-chip coupling loss, despite a higher gain coefficient with 980 nm pumping. The C-band signal was combined with the 1480-nm forward pump light by a 1480/1550-nm WDM coupler, and the backward pump light was coupled from the output side of the EDWA via another 1480/1550-nm coupler. The remaining forward pump light was filtered out by the WDM coupler. Fiber polarization controllers (FPCs) were used to align the polarizations of both the signal and the pump lights to the TE mode of the EDWA. The EDWA temperature was set to 25°C by a thermo-electric cooler placed under the chip.

**Setup for the channel under test (CUT).** The 170-GBaud electrical signal was generated by a 2-channel 256-GSa/s arbitrary waveform generator (AWG) having an analog bandwidth of about 80 GHz. A radio-frequency (RF) driver was embedded in each AWG channel to provide an output peak-to-peak voltage of 2.5 V. The AWG outputs drove a single-polarization $LiNbO_3$ I/Q Mach-Zehnder modulator (MZM) with a 3-dB bandwidth of about 35 GHz and a smooth frequency response decay extended to over 80 GHz. The source laser was an ITLA tunable across the full C-band with <100-kHz linewidth. Since the 35-GHz modulator had a high modulation loss due to its insufficient bandwidth, an EDFA was inserted ahead of the modulator to adjust its output power to a common level, which is not necessary for a practical 1.6T-class coherent modulator. After the EDWA amplified the signal-polarization signal, the PDM signal was emulated by polarization-combining the single-polarization signal with its decorrelated copy after 10-m fiber delay line (*i.e.*, about 50 ns). Because the coherence length of the optical ASE noise was much shorter than that of the 170-GBaud signal, the ASE noise of the EDWA was simultaneously decorrelated on the two polarizations. Therefore, our measured transmission performance is equivalent to that in a DP EDWA-based system. The signal was either back-to-back (B2B) tested or transmitted over 10-km SSMF (attenuation of 0.2 dB/km), and then intradyne-detected by a standard DP coherent receiver consisting of a 90-degree optical hybrid and four balanced photodiodes (BPDs) with a 3-dB bandwidth of about 100 GHz. The BPD outputs were digitized by a 4-channel 256-GSa/s real-time oscilloscope (RTO) with an analog bandwidth of 113 GHz, and then processed by offline DSP.

**Setup for the WDM transmission.** The 16-channel WDM signal consisted of one CUT and 15 loading channels. These side channels used either ITLA or distributed feedback (DFB) laser diodes on a 200-GHz WDM grid as the light sources, whose output power was adjusted to achieve flat WDM spectrum after the EDWA, as shown in Fig. 2c. A second I/Q modulator was driven by a pair of AWG outputs to modulate the 15 side wavelengths with the same 170-GBaud I/Q signal, followed by 1-km SSMF spool to temporally decorrelate these channels. The chromatic dispersion of 1-km fiber induced about 4.6 symbols (at 170 GBaud) of delay between neighboring channels prior to transmission. The CUT and 15 side channels were combined by a 3-dB coupler with an aligned polarization. Because the EDWA performance was measured by the CUT, we did not limit the usage of EDFAs for WDM loading channels. In the setup for WDM loads, the 1[st] EDFA compensated for the 14-dB loss from the wavelength-nonselective WDM multiplexer, and the 2[nd] EDFA compensated for the modulation loss, whose gain was adjusted to equalize the power of 15 side channels with respect to the CUT. The input power to the EDWA was set to -8 dBm per channel (off chip) and 4 dBm in total for 16 channels. The 16-channel WDM signal was transmitted over 81-km Corning Vascade® EX2000 ultra-low attenuation (0.16 dB/km) fiber, with a total span loss of about 14 dB including splicing and connector loss. At the receiver, the WDM signal was pre-amplified by a single-mode EDFA, and then demultiplexed by a WSS with 200-GHz passband bandwidth. The CUT was tested by the same coherent receiver as in the single-channel setup.

**Modulation format and offline DSP.** Each WDM channel was modulated by a 170-GBd PS [19] 64-QAM signal with an entropy of 5.7 bits/symbol. The signal was pulse-shaped at two samples per symbol by an RRC filter with a roll-off factor of 0.1, and then resampled to 256 GSa/s. After linear pre-equalization and I/Q imbalance compensation, the I/Q signals were quantized and loaded to the AWG memory for D/A conversions. The receiver offline DSP included frontend corrections (i.e., frequency response compensation of receiver components, I/Q orthogonalization), chromatic dispersion compensation, matched filtering, 4×4 widely linear multi-input-multi-output (MIMO) equalization embedded with carrier recovery, nonlinear post-equalization and NGMI evaluation. The I/Q modulator introduced a strong frequency-dependent I/Q imbalance at the transmitter as its 3-dB bandwidth was much smaller than that of the 170-GBaud signal. Therefore, a widely linear MIMO with an extended dimension from 2×2 to 4×4 was applied to demultiplex the PDM signal and correct the transmitter I/Q imbalance simultaneously. The MIMO equalizer outputs were four real-valued data streams corresponding to two I/Q pairs of the transmitted PDM signal, which were further processed by four nonlinear post-equalizers in parallel, mainly to compensate for the RF driver nonlinearity as well as the bandwidth limit of the electro-optic transmitter.

**System metric and net rate evaluation.** For SD-FEC, NGMI has been a widely accepted metric to predict the post-FEC performance. To claim the net bit rate, we select a concatenated FEC scheme with an inner spatially coupled low-density parity-check (SC-LDPC) code (rate of 0.8469) [22] and an outer hard-decision BCH (8191,8126,5) code. The net bit rate is calculated as $R = H - (1 - c) \log_2 M$, where $H$=5.7 bits/symbol is the source entropy, $c$=0.8402 is the concatenated code rate and $M$=64 is the QAM order. For the PS-QAM signal, the net bit rate is 4.74 bits/symbol, and the total net rate of a PDM 170-GBaud channel is 1.612 Tb/s.

**Figure 1**

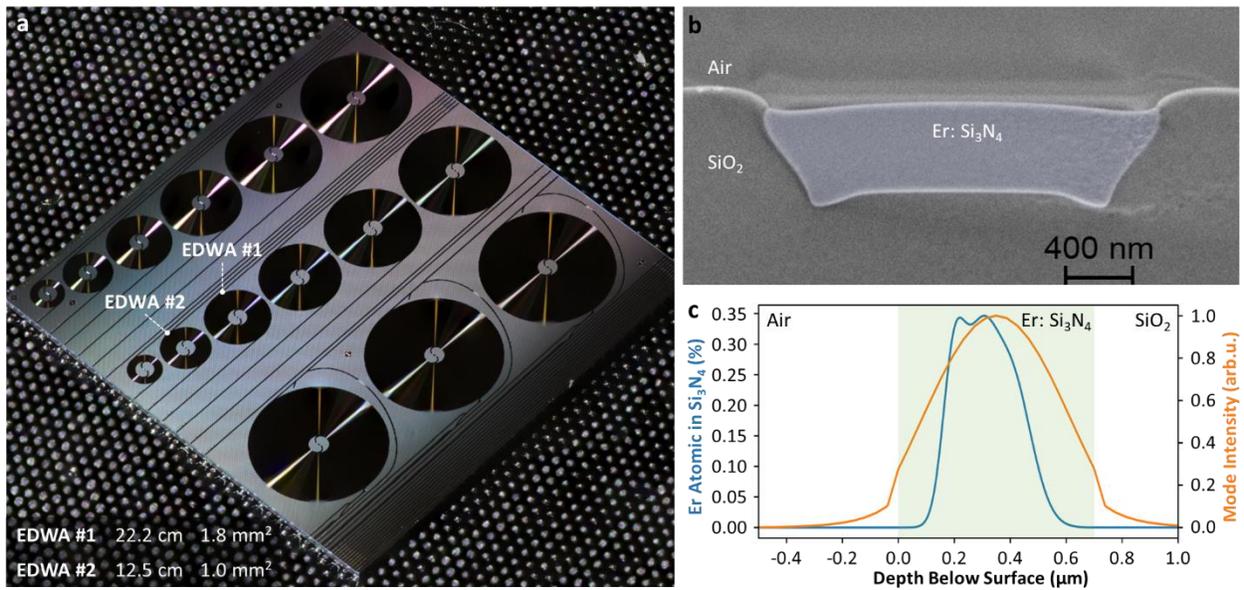

**Figure 1 | Erbium-doped waveguide amplifier (EDWA) on a silicon nitride (Si$_3$N$_4$) photonic platform.** (a) Top-view photographs of a Si$_3$N$_4$ chip hosting multiple EDWAs. (b) False-colored scanning electron microscopy (SEM) image of the waveguide cross-section of the ion-implanted gain waveguide. (c) Simulated mode field intensity profile integrated in the horizontal direction and the erbium doping concentration as a function of depth.

**Figure 2**

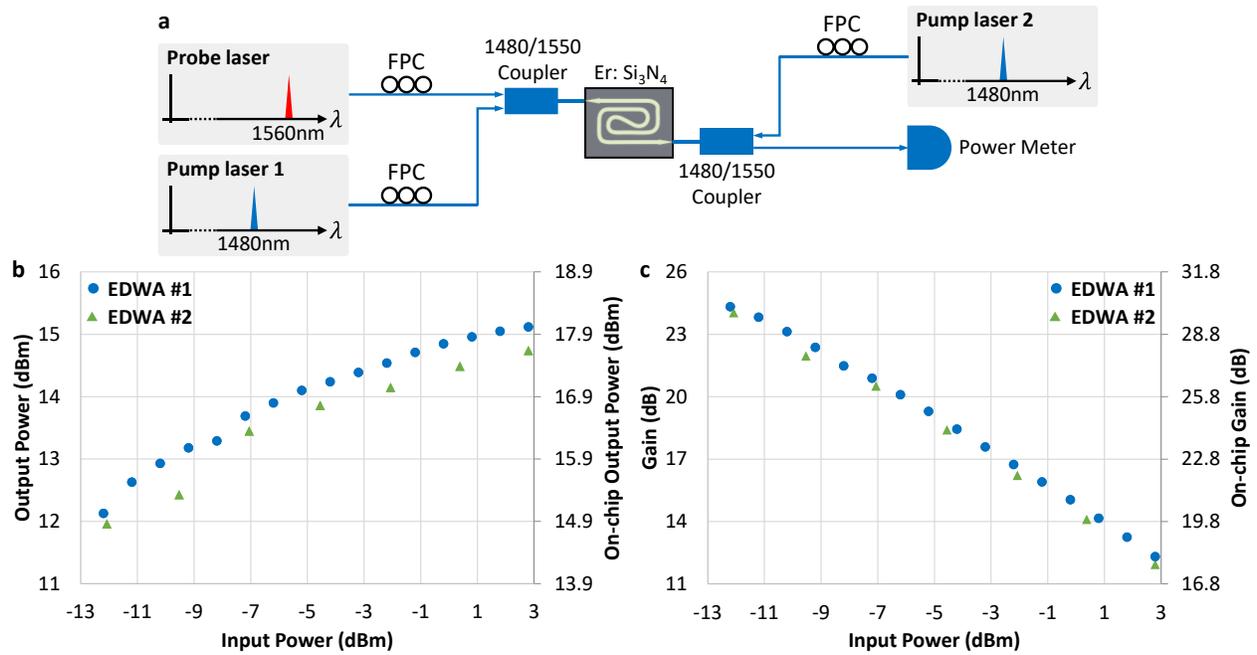

**Figure 2 | Experimental characterization of the EDWAs.** (a) Experimental setup to measure the output optical power and the gain of the EDWA. (b) Output optical power versus input optical power and (c) gain versus input optical power for EDWAs #1 and #2 in Fig. 1 with length of 22.2 cm and 12.5 cm, respectively. The secondary axes show the "on-chip" output power and "on-chip" output gain, which have de-embedded the fiber-to-chip coupling loss (about 2.9 dB/facet).

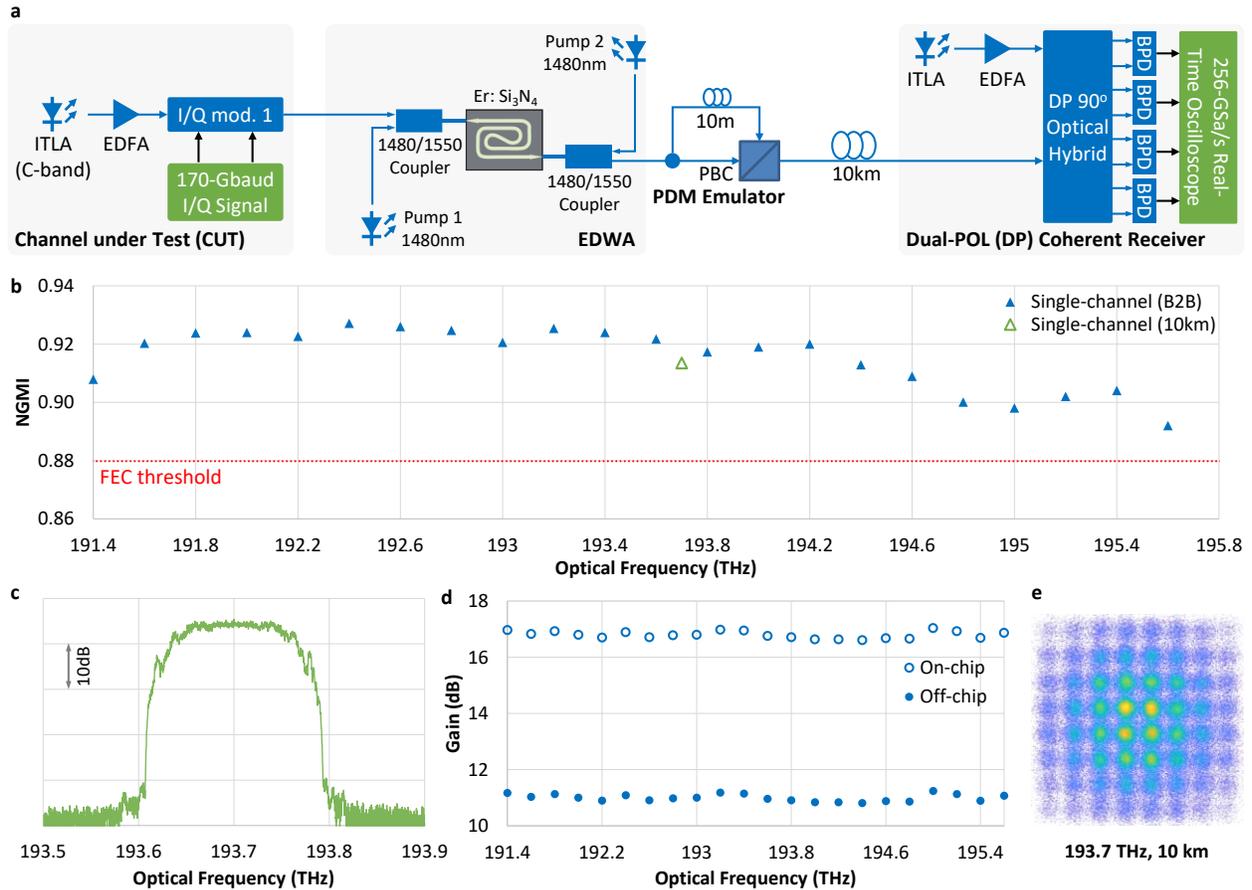

**Figure 3 | Single-channel 1.6-Tb/s transmission (Experiment 1). a.** Transmission system setup. The EDWA used in this measurement was device #1 in Fig. 1a with a spiral length of 22.2 cm. The integrated tunable laser assembly (ITLA) changed its emitting frequency from 191.4 to 195.6 THz with a step size of 200 GHz. **b.** Measured NGMI performance of dual-polarization (DP) 170-GBaud PS 64-QAM (5.7-bits/symbol entropy) signals at different optical frequencies. The signal was tested in back-to-back (B2B) condition or transmitted through 10-km SSMF. The red dashed line indicates the NGMI threshold (0.8798) of a rate-0.8402 SD-FEC code. **c.** Optical spectrum of a 170-GBaud signal (with a pulse shaping roll-off of 0.1) at the output of the I/Q modulator (ahead of the EDWA). **d.** Optical amplification gain at each frequency. The off-chip gain was about 5.8 dB lower than the on-chip gain due to the fiber coupling loss of 2.9 dB/facet. **e.** A received constellation of the PS 64-QAM signal at 193.7 THz, after 10-km transmission.

**Figure 4**

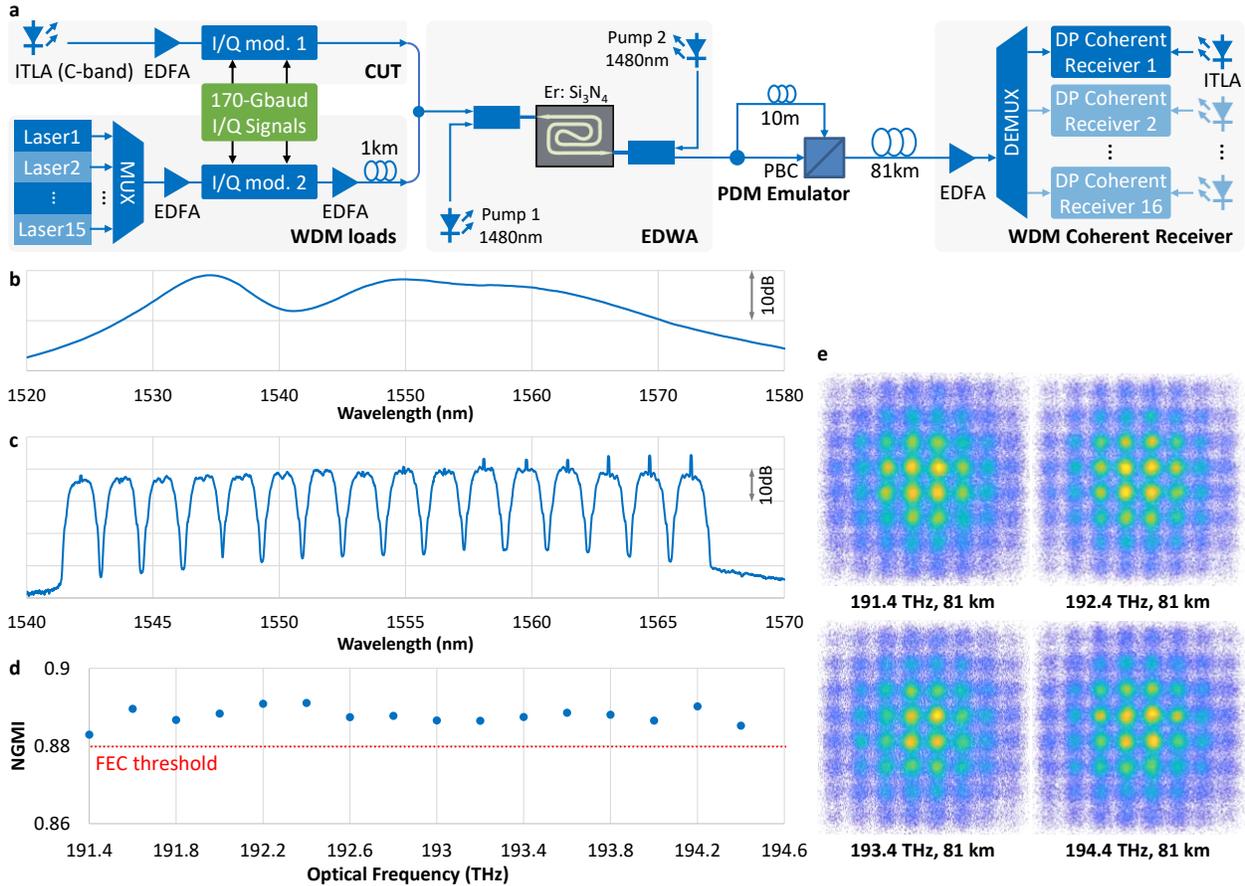

**Figure 4 | 25.6-Tb/s transmission with 16 WDM channels (Experiment 2). a.** Transmission system setup. The 15 WDM loading channels were generated by sending 15 wavelengths to one I/Q modulator and then decorrelated by the chromatic dispersion accumulated over 1-km SSMF. Using one DP coherent receiver with a tunable local oscillator, the WDM channels were measured one at a time after wavelength demultiplexing. The EDWA used in this measurement was device #2 in Fig. 1a with a spiral length of 12.5 cm. The input power to the EDWA was 4 dBm and the fiber-coupled output power was 14 dBm (i.e., 10-dB off-chip gain). **b.** Optical ASE noise spectrum of the EDWA without signal inputs. **c.** Optical spectrum of the WDM signal after being amplified by the EDWA. **d.** Measured NGMI performance for the 16 WDM channels after 81-km Corning Vascade® EX2000 fiber (attenuation of 0.16 dB/km). The red dashed line is the NGMI threshold (0.8798) of a rate-0.8402 SD-FEC code. **e.** Example constellations of 170-GBaud PS 64-QAM signals at frequencies of 191.4, 192.4, 193.4 and 194.4 THz, respectively.